
\font\twelverm=cmr10 scaled 1200    \font\twelvei=cmmi10 scaled 1200
\font\twelvesy=cmsy10 scaled 1200   \font\twelveex=cmex10 scaled 1200
\font\twelvebf=cmbx10 scaled 1200   \font\twelvesl=cmsl10 scaled 1200
\font\twelvett=cmtt10 scaled 1200   \font\twelveit=cmti10 scaled 1200
\skewchar\twelvei='177   \skewchar\twelvesy='60
\def\twelvepoint{\normalbaselineskip=12.4pt
  \abovedisplayskip 12.4pt plus 3pt minus 9pt
  \belowdisplayskip 12.4pt plus 3pt minus 9pt
  \abovedisplayshortskip 0pt plus 3pt
  \belowdisplayshortskip 7.2pt plus 3pt minus 4pt
  \smallskipamount=3.6pt plus1.2pt minus1.2pt
  \medskipamount=7.2pt plus2.4pt minus2.4pt
  \bigskipamount=14.4pt plus4.8pt minus4.8pt
  \def\rm{\fam0\twelverm}          \def\it{\fam\itfam\twelveit}%
  \def\sl{\fam\slfam\twelvesl}     \def\bf{\fam\bffam\twelvebf}%
  \def\mit{\fam 1}                 \def\cal{\fam 2}%
  \def\tt{\twelvett}
  \textfont0=\twelverm   \scriptfont0=\tenrm   \scriptscriptfont0=\sevenrm
  \textfont1=\twelvei    \scriptfont1=\teni    \scriptscriptfont1=\seveni
  \textfont2=\twelvesy   \scriptfont2=\tensy   \scriptscriptfont2=\sevensy
  \textfont3=\twelveex   \scriptfont3=\twelveex  \scriptscriptfont3=\twelveex
  \textfont\itfam=\twelveit
  \textfont\slfam=\twelvesl
  \textfont\bffam=\twelvebf \scriptfont\bffam=\tenbf
  \scriptscriptfont\bffam=\sevenbf
  \normalbaselines\rm}

\def\beginlinemode{\endmode
  \begingroup\parskip=0pt \obeylines\def\\{\par}\def\endmode{\par\endgroup}}
\def\beginparmode{\endmode
  \begingroup \def\endmode{\par\endgroup}}
\let\endmode=\par
{\obeylines\gdef\
{}}
\def\singlespace{\baselineskip=\normalbaselineskip}
\def\oneandahalfspace{\baselineskip=\normalbaselineskip
  \multiply\baselineskip by 3 \divide\baselineskip by 2}
\def\doublespace{\baselineskip=\normalbaselineskip \multiply\baselineskip by 2}
\newcount\firstpageno
\firstpageno=2
\footline={\ifnum\pageno<\firstpageno{\hfil}\else{\hfil\twelverm\folio\hfil}\fi}
\let\rawfootnote=\footnote              
\def\footnote#1#2{{\rm\singlespace\parindent=0pt\rawfootnote{#1}{#2}}}
\def\raggedcenter{\leftskip=2em plus 12em \rightskip=\leftskip
  \parindent=0pt \parfillskip=0pt \spaceskip=.3333em \xspaceskip=.5em
  \pretolerance=9999 \tolerance=9999
  \hyphenpenalty=9999 \exhyphenpenalty=9999 }
\parskip=\medskipamount
\twelvepoint            
\overfullrule=0pt       
\def\preprintno#1{
 \rightline{\rm #1}}    
\def\author                     
  {\vskip 3pt plus 0.2fill \beginlinemode
   \singlespace \raggedcenter \twelvesc}
\def\affil                      
  {\vskip 3pt plus 0.1fill \beginlinemode
   \oneandahalfspace \raggedcenter \sl}
\def\abstract                   
  {\vskip 3pt plus 0.3fill \beginparmode
   \doublespace \narrower \noindent ABSTRACT: }
\def\endtitlepage               
  {\endpage                     
   \body}
\def\body                       
  {\beginparmode}               

\def\subhead#1{                 
  \vskip 0.25truein             
  {\raggedcenter #1 \par}
   \nobreak\vskip 0.1truein\nobreak}
\def\refto#1{$|{#1}$}           
\def\references                 
  {\subhead{References}         
   \beginparmode
   \frenchspacing \parindent=0pt \leftskip=1truecm
   \parskip=8pt plus 3pt \everypar{\hangindent=\parindent}}
\gdef\refis#1{\indent\hbox to 0pt{\hss#1.~}}    
\gdef\journal#1, #2, #3, 1#4#5#6{               
    {\sl #1~}{\bf #2}, #3, (1#4#5#6)}           
\def\refstylenp{                
  \gdef\refto##1{ [##1]}                                
  \gdef\refis##1{\indent\hbox to 0pt{\hss##1)~}}        
  \gdef\journal##1, ##2, ##3, ##4 {                     
     {\sl ##1~}{\bf ##2~}(##3) ##4 }}
\def\refstyleprnp{              
  \gdef\refto##1{ [##1]}                                
  \gdef\refis##1{\indent\hbox to 0pt{\hss##1)~}}        
  \gdef\journal##1, ##2, ##3, 1##4##5##6{               
    {\sl ##1~}{\bf ##2~}(1##4##5##6) ##3}}
\def\pr{\journal Phys. Rev., }
\def\prd{\journal Phys. Rev. D, }
\def\prl{\journal Phys. Rev. Lett., }
\def\prpts{\journal Phys. Rep., }
\def\np{\journal Nucl. Phys., }
\def\pl{\journal Phys. Lett., }

\def\endreferences{\body}
\def\endpage                    
  {\vfill\eject}
\def\endpaper                   
  {\endmode\vfill\supereject}
\def\endit
  {\endpaper\end}
\def\ref#1{Ref. #1}                     
\def\Ref#1{Ref. #1}                     

\def\m@th{\mathsurround=0pt }
\font\twelvesc=cmcsc10 scaled 1200
\def\cite#1{{#1}}
\def\(#1){(\call{#1})}
\def\call#1{{#1}}
\def\taghead#1{}
\def\leaderfill{\leaders\hbox to 1em{\hss.\hss}\hfill}
\def\twiddle{\lower.9ex\rlap{$\kern-.1em\scriptstyle\sim$}}
\def\bigtwiddle{\lower1.ex\rlap{$\sim$}}
\def\gtwid{\mathrel{\raise.3ex\hbox{$>$\kern-.75em\lower1ex\hbox{$\sim$}}}}
\def\ltwid{\mathrel{\raise.3ex\hbox{$<$\kern-.75em\lower1ex\hbox{$\sim$}}}}
\def\square{\kern1pt\vbox{\hrule height 1.2pt\hbox{\vrule width 1.2pt\hskip 3pt
   \vbox{\vskip 6pt}\hskip 3pt\vrule width 0.6pt}\hrule height 0.6pt}\kern1pt}
\catcode`@=11
\newcount\tagnumber\tagnumber=0

\immediate\newwrite\eqnfile
\newif\if@qnfile\@qnfilefalse
\def\write@qn#1{}
\def\writenew@qn#1{}
\def\w@rnwrite#1{\write@qn{#1}\message{#1}}
\def\@rrwrite#1{\write@qn{#1}\errmessage{#1}}

\def\taghead#1{\gdef\t@ghead{#1}\global\tagnumber=0}
\def\t@ghead{}

\expandafter\def\csname @qnnum-3\endcsname
  {{\t@ghead\advance\tagnumber by -3\relax\number\tagnumber}}
\expandafter\def\csname @qnnum-2\endcsname
  {{\t@ghead\advance\tagnumber by -2\relax\number\tagnumber}}
\expandafter\def\csname @qnnum-1\endcsname
  {{\t@ghead\advance\tagnumber by -1\relax\number\tagnumber}}
\expandafter\def\csname @qnnum0\endcsname
  {\t@ghead\number\tagnumber}
\expandafter\def\csname @qnnum+1\endcsname
  {{\t@ghead\advance\tagnumber by 1\relax\number\tagnumber}}
\expandafter\def\csname @qnnum+2\endcsname
  {{\t@ghead\advance\tagnumber by 2\relax\number\tagnumber}}
\expandafter\def\csname @qnnum+3\endcsname
  {{\t@ghead\advance\tagnumber by 3\relax\number\tagnumber}}

\def\equationfile{%
  \@qnfiletrue\immediate\openout\eqnfile=\jobname.eqn%
  \def\write@qn##1{\if@qnfile\immediate\write\eqnfile{##1}\fi}
  \def\writenew@qn##1{\if@qnfile\immediate\write\eqnfile
    {\noexpand\tag{##1} = (\t@ghead\number\tagnumber)}\fi}
}

\def\callall#1{\xdef#1##1{#1{\noexpand\call{##1}}}}
\def\call#1{\each@rg\callr@nge{#1}}

\def\each@rg#1#2{{\let\thecsname=#1\expandafter\first@rg#2,\end,}}
\def\first@rg#1,{\thecsname{#1}\apply@rg}
\def\apply@rg#1,{\ifx\end#1\let\next=\relax%
\else,\thecsname{#1}\let\next=\apply@rg\fi\next}

\def\callr@nge#1{\calldor@nge#1-\end-}
\def\callr@ngeat#1\end-{#1}
\def\calldor@nge#1-#2-{\ifx\end#2\@qneatspace#1 %
  \else\calll@@p{#1}{#2}\callr@ngeat\fi}
\def\calll@@p#1#2{\ifnum#1>#2{\@rrwrite{Equation range #1-#2\space is bad.}
\errhelp{If you call a series of equations by the notation M-N, then M and
N must be integers, and N must be greater than or equal to M.}}\else%
 {\count0=#1\count1=#2\advance\count1
by1\relax\expandafter\@qncall\the\count0,%
  \loop\advance\count0 by1\relax%
    \ifnum\count0<\count1,\expandafter\@qncall\the\count0,%
  \repeat}\fi}

\def\@qneatspace#1#2 {\@qncall#1#2,}
\def\@qncall#1,{\ifunc@lled{#1}{\def\next{#1}\ifx\next\empty\else
  \w@rnwrite{Equation number \noexpand\(>>#1<<) has not been defined yet.}
  >>#1<<\fi}\else\csname @qnnum#1\endcsname\fi}

\let\eqnono=\eqno
\def\eqno(#1){\tag#1}
\def\tag#1$${\eqnono(\displayt@g#1 )$$}

\def\aligntag#1\endaligntag
  $${\gdef\tag##1\\{&(##1 )\cr}\eqalignno{#1\\}$$
  \gdef\tag##1$${\eqnono(\displayt@g##1 )$$}}

\def\eqalignno#1{\displ@y \tabskip\centering
  \halign to\displaywidth{\hfil$\displaystyle{##}$\tabskip\z@skip
    &$\displaystyle{{}##}$\hfil\tabskip\centering
    &\llap{$\displayt@gpar##$}\tabskip\z@skip\crcr
    #1\crcr}}

\def\displayt@gpar(#1){(\displayt@g#1 )}

\def\displayt@g#1 {\rm\ifunc@lled{#1}\global\advance\tagnumber by1
        {\def\next{#1}\ifx\next\empty\else\expandafter
        \xdef\csname @qnnum#1\endcsname{\t@ghead\number\tagnumber}\fi}%
  \writenew@qn{#1}\t@ghead\number\tagnumber\else
        {\edef\next{\t@ghead\number\tagnumber}%
        \expandafter\ifx\csname @qnnum#1\endcsname\next\else
        \w@rnwrite{Equation \noexpand\tag{#1} is a duplicate number.}\fi}%
  \csname @qnnum#1\endcsname\fi}

\def\ifunc@lled#1{\expandafter\ifx\csname @qnnum#1\endcsname\relax}

\let\@qnend=\end\gdef\end{\if@qnfile
\immediate\write16{Equation numbers written on []\jobname.EQN.}\fi\@qnend}

\catcode`@=12
\refstyleprnp
\catcode`@=11
\newcount\r@fcount \r@fcount=0
\def\refreset{\global\r@fcount=0}
\newcount\r@fcurr
\immediate\newwrite\reffile
\newif\ifr@ffile\r@ffilefalse
\def\w@rnwrite#1{\ifr@ffile\immediate\write\reffile{#1}\fi\message{#1}}

\def\writer@f#1>>{}
\def\referencefile{
  \r@ffiletrue\immediate\openout\reffile=\jobname.ref%
  \def\writer@f##1>>{\ifr@ffile\immediate\write\reffile%
    {\noexpand\refis{##1} = \csname r@fnum##1\endcsname = %
     \expandafter\expandafter\expandafter\strip@t\expandafter%
     \meaning\csname r@ftext\csname r@fnum##1\endcsname\endcsname}\fi}%
  \def\strip@t##1>>{}}

\def\citeall#1{\xdef#1##1{#1{\noexpand\cite{##1}}}}
\def\cite#1{\each@rg\citer@nge{#1}}	

\def\each@rg#1#2{{\let\thecsname=#1\expandafter\first@rg#2,\end,}}
\def\first@rg#1,{\thecsname{#1}\apply@rg}	
\def\apply@rg#1,{\ifx\end#1\let\next=\relax
\else,\thecsname{#1}\let\next=\apply@rg\fi\next}

\def\citer@nge#1{\citedor@nge#1-\end-}	
\def\citer@ngeat#1\end-{#1}
\def\citedor@nge#1-#2-{\ifx\end#2\r@featspace#1 
  \else\citel@@p{#1}{#2}\citer@ngeat\fi}	
\def\citel@@p#1#2{\ifnum#1>#2{\errmessage{Reference range #1-#2\space is bad.}%
    \errhelp{If you cite a series of references by the notation M-N, then M and
    N must be integers, and N must be greater than or equal to M.}}\else%
 {\count0=#1\count1=#2\advance\count1 by1\relax\expandafter\r@fcite\the\count0,
  \loop\advance\count0 by1\relax
    \ifnum\count0<\count1,\expandafter\r@fcite\the\count0,%
  \repeat}\fi}

\def\r@featspace#1#2 {\r@fcite#1#2,}	
\def\r@fcite#1,{\ifuncit@d{#1}
    \newr@f{#1}%
    \expandafter\gdef\csname r@ftext\number\r@fcount\endcsname%
                     {\message{Reference #1 to be supplied.}%
                      \writer@f#1>>#1 to be supplied.\par}%
 \fi%
 \csname r@fnum#1\endcsname}
\def\ifuncit@d#1{\expandafter\ifx\csname r@fnum#1\endcsname\relax}%
\def\newr@f#1{\global\advance\r@fcount by1%
    \expandafter\xdef\csname r@fnum#1\endcsname{\number\r@fcount}}

\let\r@fis=\refis			
\def\refis#1#2#3\par{\ifuncit@d{#1}
   \newr@f{#1}%
   \w@rnwrite{Reference #1=\number\r@fcount\space is not cited up to now.}\fi%
  \expandafter\gdef\csname r@ftext\csname r@fnum#1\endcsname\endcsname%
  {\writer@f#1>>#2#3\par}}

\def\ignoreuncited{
   \def\refis##1##2##3\par{\ifuncit@d{##1}%
     \else\expandafter\gdef\csname r@ftext\csname
r@fnum##1\endcsname\endcsname%
     {\writer@f##1>>##2##3\par}\fi}}

\def\r@ferr{\endreferences\errmessage{I was expecting to see
\noexpand\endreferences before now;  I have inserted it here.}}
\let\r@ferences=\references
\def\references{\r@ferences\def\endmode{\r@ferr\par\endgroup}}

\let\endr@ferences=\endreferences
\def\endreferences{\r@fcurr=0
  {\loop\ifnum\r@fcurr<\r@fcount
    \advance\r@fcurr by 1\relax\expandafter\r@fis\expandafter{\number\r@fcurr}%
    \csname r@ftext\number\r@fcurr\endcsname%
  \repeat}\gdef\r@ferr{}\global\r@fcount=0\endr@ferences}

\let\r@fend=\endpaper\gdef\endpaper{\ifr@ffile
\immediate\write16{Cross References written on []\jobname.REF.}\fi\r@fend}

\catcode`@=12

\citeall\refto		
\citeall\ref		%
\citeall\Ref		%

\referencefile

\def\ssc{\scriptscriptstyle}

\def\frac#1/#2{#1 / #2}

\def\neuphys{Department of Physics\\Northeastern University\\Boston MA 02115}
\def\ssc{Physics Research Division\\Superconducting Super Collider Laboratory
\\Dallas TX 75237}

\def\oneandthreefifthsspace{\baselineskip=\normalbaselineskip
  \multiply\baselineskip by 8 \divide\baselineskip by 5}

\font\titlefont=cmr10 scaled\magstep3
\def\bigtitle                      
  {\null\vskip 3pt plus 0.2fill
   \beginlinemode \doublespace \raggedcenter \titlefont}

\def\sul{\tilde u_L}
\def\sur{\tilde u_R}
\def\sdl{\tilde d_L}
\def\sdr{\tilde d_R}
\def\ssl{\tilde s_L}
\def\ssr{\tilde s_R}
\def\scl{\tilde c_L}
\def\scr{\tilde c_R}
\def\sbl{\tilde b_L}
\def\sbr{\tilde b_R}

\def\sel{\tilde e_L}
\def\ser{\tilde e_R}
\def\smul{\tilde \mu_L}
\def\smur{\tilde \mu_R}

\def\seln{\tilde \nu_e}
\def\smun{\tilde \nu_\mu}

\def\mmtone{M_{\tilde t_1}}
\def\mmttwo{M_{\tilde t_2}}
\def\mmbl{M_{\tilde b_L}}

\def\mmul{M_{\tilde u_L}}
\def\mmdl{M_{\tilde d_L}}
\def\mmur{M_{\tilde u_R}}
\def\mmdr{M_{\tilde d_R}}
\def\mmel{M_{\tilde e_L}}

\def\mmnl{M_{\tilde \nu_e}}
\def\mmer{M_{\tilde e_R}}

\def\mbl{m_{\tilde b_L}}
\def\mbr{m_{\tilde b_R}}
\def\mul{m_{\tilde u_L}}
\def\mdl{m_{\tilde d_L}}
\def\mur{m_{\tilde u_R}}
\def\mdr{m_{\tilde d_R}}

\def\st{\sin^2 \theta_W}
\def\mglu{M_{\tilde g}}

\oneandthreefifthsspace

\preprintno{NUB-3067-93TH}
\preprintno{UFIFT-HEP-93-16}
\preprintno{SSCL-Preprint-439}
\preprintno{June, 1993}
\bigtitle{Sparticle Spectrum Constraints}
\bigskip
\author Stephen P. Martin
\affil\neuphys
\centerline{and}
\author Pierre Ramond$^\dagger $
\affil\ssc
\body

\footnote{}{$^\dagger$ Permanent address: Institute for
Fundamental Theory, Physics Department, University of Florida,
Gainesville FL 32611}

\abstract
The supersymmetric standard model with supergravity-inspired soft
breaking terms predicts a rich pectrum of sparticles to be discovered
at the SSC, LHC and NLC. Because there are more supersymmetric particles
than unknown parameters, one can write down sum rules relating their masses.
We discuss the pectrum of sparticles from this point of view.
Some of the sum rules do not depend on the input parameters and can be used to
test the consistency of the model, while others are useful in determining
the input parameters of the theory. If supersymmetry is discovered but
the sum rules turn out to be violated, it will be evidence of new physics
beyond the minimal supersymmetric standard model with universal soft
supersymmetry-breaking terms.

\endtitlepage

\subhead{1. INTRODUCTION}
\taghead{1.}

The extension of the Standard Model
to $N=1$ supersymmetry[\cite{reviews}]
is totally straightforward outside of the Higgs
sector: to every chiral fermion (quark or lepton) one associates
a complex spinless boson (squark or slepton), and to each gauge field
(gluon, $W$-boson, $Z$-boson, photon), one adds a spin $1/2$ Majorana gaugino
(gluino, wino, zino, photino).
The extension of the Higgs sector to supersymmetry is
less straightforward, since by extending the single Higgs doublet of the
Standard Model to a chiral Higgsino, one induces local (ABJ) and
global (Witten) anomalies.
This is easily solved by adding its vector-like completion; the result is a
theory with two Higgs doublets of opposite weak hypercharge. This is
fortuitous since the nature of supersymmetric couplings itself requires
two Higgs doublets if both up and down type quarks and leptons are
to be massive. However, theories with two Higgs doublets show an
additional chiral global symmetry, of the type introduced by
Peccei and Quinn. This symmetry is broken explicitly by QCD instanton
effects and spontaneously by the electroweak symmetry breaking.
This results in a pseudo-Nambu-Goldstone boson, the axion, of
the type ruled out by experiment.

To be in accord with experiment, the PQ symmetry must be broken.
Fortunately, this can be done without introducing new fields in the
theory by adding a term which preserves supersymmetry; it has dimension
3, and is parametrized by a coupling with dimension of mass called
$\mu$. Its numerical value is to be considered one of the parameters of
the $N=1$ Standard Model. The model also comes with a potential
but it is not capable of breaking the electroweak symmetry, since
the Higgs scalar bosons have the same positive mass squared, $\mu^2$.

In order to bring this model closer to reality, one must break supersymmetry.
In the absence of any concrete theory of supersymmetry breaking,
the effect is mocked up in the low energy Lagrangian by including terms
which break the supersymmetry softly while preserving the gauge symmetries.

The generalization of the model to
include supergravity allows for such a mechanism of
supersymmetry breaking, with a particular set of soft terms specified
at a given input scale[\cite{supergrav}]. They are:
\item{-} masses for the three gauginos [$M_1$ for weak hypercharge,
$M_2$ for $SU(2)_L$ and $M_3$ for $SU(3)_c$];
\item{-} a common mass $m_0$ for all the spinless particles
in the theory (squark, slepton, Higgs);
\item{-} cubic interactions among the squarks, sleptons and Higgs
as allowed by $R$-parity and
the gauge interactions of the theory, each equal at the input scale to the
corresponding Yukawa coupling multiplied by a universal parameter
$A$; and
\item{-} a scalar (mass)$^2$ term in the Higgs sector
which breaks both the Peccei-Quinn symmetry and supersymmetry.

In this scheme, the supersymmetry-breaking sector is parametrized by six
masses; three gaugino masses $M_i$, the common scalar mass $m_0$,
the trilinear scalar coupling parameter $A$, and the PQ-breaking and
supersymmetry-breaking parameter $B$. Since the gauginos have not
been observed to date, the masses $M_i$ must certainly be non-zero.
In fact, since they are strictly multiplicatively renormalized,
they must not vanish at any scale for which the renormalization
group equations are valid. It is possible that each of the other
supersymmetry-breaking parameters are zero at the input scale, although this
will not be maintained under renormalization group evolution.
The low energy values of all the soft breaking parameters are constrained
by the fact that no sparticles have been found yet.

A remarkable feature of the theory is that, with the supersymmetry
breaking parameters specified at some high scale $M_X$, it is possible to
trigger electroweak symmetry breaking[\cite{trigger}].
It is even more amazing that the present
bounds on the top-quark mass, which is constrained  (in the context of the
Standard Model) to be between 120 and 200 GeV by experiment,
yields the correct value of $M_Z$ for values of the supersymmetry breaking
parameters not far above their experimental lower
limits. One of the consequences of
such a picture is that the superpartners of the elementary particles
would have masses in the hundreds of GeV, quite accessible to the
next generation of colliders: SSC, LHC and NLC[\cite{JLC}].
Another remarkable consequence of the mechanism is that it suggests
that the three gauge couplings of the Standard Model
have a common origin[\cite{unification}]
around $10^{15}-10^{16}$ GeV, providing a strong hint in favor of a Grand
Unified Theory (GUT)[\cite{GUT}].
This in turn dovetails nicely with the requirement of
$R$-parity conservation, which is necessary in order to avoid superfast
proton decay, and which arises most naturally in supersymmetric
GUTs with gauged $B-L$[\cite{rparity},\cite{sscfgrp}].

The scale at which the supersymmetry-breaking parameters are specified
is in principle undetermined as long as it is below the Planck
mass. However, if it is too far below, the magnitude of the electroweak
breaking comes out too low, given the lower bound on the top-quark mass.
Thus a remarkably consistent picture emerges with the supersymmetry breaking
parameters specified at $10^{15}-10^{16}$ GeV, the scale at which the gauge
couplings unify. Another coincidence is that around that scale, the relation
$m_b = m_\tau$ is favored[\cite{btau}] by the data.
One of the mysteries is that the PQ-breaking parameter $\mu$ is constrained
to be of the same order of magnitude as the supersymmetry-breaking
parameters. Since $\mu$ is logically uncorrelated with
supersymmetry-breaking from the low-energy point of view, this hints at
a deeper mechanism which would link PQ and supersymmetry breakings.

The masses of the
superpartners are determined in terms of the soft breaking
parameters[\cite{spartnersa},\cite{spartnersb}].
Various
authors[\cite{num},\cite{AN},\cite{AN2},\cite{numer},\cite{nume},\cite{sla}]
have presented numerical
results based on computer analysis for the sparticle masses, using different
sets of input  parameters. Since there are more superpartners than
breaking parameters,
there are many sum rules among superpartner masses. These sum
rules will test the validity of this picture, and will be of importance
for the SSC, LHC, and NLC. It is the purpose of this paper to
present these sum rules in a simple, unified format, without using
computers. Some are new;
some have already appeared in the literature, but not all in one place.
Their study will enable us to offer some specific scenarios in conducting
the experimental search for superpartners.

Since the superpartner masses are in the several hundred GeV range,
we neglect the masses and Yukawa couplings of the leptons and quarks of the
first two families. It follows that we need only consider the
Yukawa couplings $y_t$, $y_b$ and $y_\tau$,
and the trilinear scalar couplings of $H_u \tilde t_L \tilde t_R^*$,
$H_d \tilde b_L \tilde b_R^*$ and $H_d \tilde \tau_L \tilde \tau_R^*$,
which we denote $y_t A_t$, $y_b A_b$ and $y_\tau A_\tau$,
respectively. At the input scale, $A_t = A_b = A_\tau$.
We denote the squarks and sleptons of the first
two families by their first-family names,
that is
${\tilde u}_L$, ${\tilde d}_L$, ${\tilde u}_R$, ${\tilde d}_R$,
${\tilde e}_L$, ${\tilde e}_R$, ${\tilde \nu}_e$.
Thus ${\tilde u}_L$ can be taken to be either the left-handed up or
charmed squark. The squarks and sleptons associated with the third family
will be denoted by
(${\tilde t}_{L,R}$, ${\tilde b}_{L,R}$, ${\tilde \tau}_{L,R}$,
${\tilde \nu}_{\tau }$).

\subhead{2. RENORMALIZATION GROUP EQUATIONS}
\taghead{2.}
In this section we remind the reader of the one-loop renormalization
group equations which are relevant to this work. They are
$$
16 \pi^2 {d  \over dt}g_i = - b_i g_i^3
\qquad\qquad i = 1,2,3;
\eqno(grge)
$$
$$
16 \pi^2 {d  \over dt} M_i = - 2b_i M_i g_i^2
\qquad\qquad i = 1,2,3;
\eqno(mgrge)
$$
for the three gauge couplings and three gaugino masses, respectively,
and above the superpartner mass thresholds,
$$
b_i = \cases{-{3\over 5} - 2 n_f & $i=1$\cr
               5 - 2 n_f & $i=2$\cr
               9 - 2 n_f & $i=3$\cr}
$$
with $i=1$ for weak hypercharge in a GUT normalization, $i=2$ for
$SU(2)_L$ and $i=3$ for $SU(3)_c$. Together, \(grge) and \(mgrge) imply
that the three quantities $M_i/\alpha_i$ do not run with scale:
$$
{M_i(t) \over \alpha_i(t)} = {M_i(t_0) \over \alpha_i(t_0)}
\> .
\eqno(mscaling)
$$

The light squark and slepton masses obey the RG equations
$$
\eqalign{
16 \pi^2 {d  \over d t} m^2_{{\tilde Q}_L} = &
-{2 \over 15} g_1^2 M_1^2
-{6 } g_2^2 M_2^2
-{32\over 3}  g_3^2 M_3^2
+ {1\over 5} g_1^2 {\rm Tr} (Ym^2)
\> , \cr
16 \pi^2 {d  \over d t} m^2_{{\tilde u}_R} = &
-{32 \over 15} g_1^2 M_1^2
-{32\over 3}  g_3^2 M_3^2
- {4\over 5} g_1^2 {\rm Tr} (Ym^2)
\> , \cr
16 \pi^2 {d  \over d t} m^2_{{\tilde d}_R} = &
-{8 \over 15} g_1^2 M_1^2
-{32\over 3}  g_3^2 M_3^2
+ {2\over 5} g_1^2 {\rm Tr} (Ym^2)
\> , \cr
16 \pi^2 {d  \over d t} m^2_{{\tilde L}_L} = &
-{6 \over 5} g_1^2 M_1^2
-6  g_2^2 M_2^2
- {3\over 5} g_1^2 {\rm Tr} (Ym^2)
\> , \cr
16 \pi^2 {d  \over d t} m^2_{{\tilde e}_R} = &
-{24 \over 5} g_1^2 M_1^2 + {6\over 5} g_1^2 {\rm Tr} (Ym^2)
\> , \cr
}\eqno(rgesqsl)
$$
where
$$
{\rm Tr}(Y m^2) = m^2_{H_u} - m^2_{H_d}
+ \sum_1^{n_f} \left ( m^2_{{\tilde Q}_L}
- 2 m^2_{{\tilde u}_R} + m^2_{{\tilde d}_R} - m^2_{{\tilde L}_L}
+ m^2_{{\tilde e}_R}
\right )
\> .
$$

The renormalization group equations for the sparticles of the third family
are different because they involve the Yukawa couplings; for the squarks
they read
$$
\eqalign{
16 \pi^2 {d  \over d t} m^2_{{\tilde t}_L, {\tilde b}_L} = &
\> 2 y_t^2 \Sigma^2_t + 2 y_b^2 \Sigma^2_b - {2 \over 15} g_1^2 M_1^2
- 6  g_2^2 M_2^2
-{32\over 3}  g_3^2 M_3^2
+ {1\over 5} g_1^2 {\rm Tr} (Ym^2)
\> , \cr
16 \pi^2 {d  \over d t} m^2_{{\tilde t}_R} = &
\> 4 y_t^2 \Sigma^2_t -{32 \over 15} g_1^2 M_1^2
-{32\over 3}  g_3^2 M_3^2
- {4\over 5} g_1^2 {\rm Tr} (Ym^2)
\> , \cr
16 \pi^2 {d  \over d t} m^2_{{\tilde b}_R} = &
\> 4 y_b^2 \Sigma^2_b -{8 \over 15} g_1^2 M_1^2
-{32\over 3}  g_3^2 M_3^2
+ {2\over 5} g_1^2 {\rm Tr} (Ym^2)
\> , \cr
}\eqno(rgestop)
$$
and for the sleptons,
$$
\eqalign{
16 \pi^2 {d  \over d t} m^2_{{\tilde \tau}_L, {\tilde \nu}_\tau} = &
\> 2 y_\tau^2 \Sigma^2_\tau  - {6 \over 5} g_1^2 M_1^2
- 6  g_2^2 M_2^2
- {3\over 5} g_1^2 {\rm Tr} (Ym^2)
\> , \cr
16 \pi^2 {d  \over d t} m^2_{{\tilde \tau}_R} = &
\> 4 y_\tau^2 \Sigma^2_\tau -{24 \over 5} g_1^2 M_1^2
+ {6\over 5} g_1^2 {\rm Tr} (Ym^2)
\> , }
\eqno(rgestau)
$$
where
$$
\eqalign{
{\Sigma}_t^2 = &(m_{H_u}^2 + m_{{\tilde t}_L}^2 + m_{{\tilde t}_R}^2 + A_t^2 )
\> ,
\cr
{\Sigma}_b^2 = &(m_{H_d}^2 + m_{{\tilde b}_L}^2 + m_{{\tilde b}_R}^2 + A_b^2 )
\> ,
\cr
{\Sigma}_\tau^2 = &(m_{H_d}^2 + m_{{\tilde \tau}_L}^2 + m_{{\tilde \tau}_R}^2
+ A_\tau^2 )
\> .
\cr
}
$$
When all of the squark, slepton and Higgs masses are the same at the
initial scale, we have
$$
{\rm Tr} (Ym^2) = m_0^2 {\rm Tr }(Y) = 0,
\eqno(trvan)
$$
as required by the absence of the gravitational mixed
anomaly. Furthermore the condition \(trvan) is maintained by the
RG evolution, and so holds at all scales. Hence we neglect it in the following,
which greatly simplifies these equations.
Fortunately, we will not need the renormalization group equations for
$m^2_{H_u}, m^2_{H_d}, A_t, A_b, A_\tau$ or $\mu$ in this analysis.

\subhead{3. FIRST AND SECOND FAMILY SQUARKS AND SLEPTONS}
\taghead{3.}

The sum rules involving masses of
the squarks and sleptons associated with the first two families
are particularly simple.
Besides the universal $m_0^2$, there are four other
contributions to the squared masses of the squarks and sleptons, as follows.

First, there are contributions from the renormalization group running
of the scalar masses down to experimental scales,
as given by \(rgesqsl).

Second, the $D^2$ term in the scalar potential
contributes to the scalar masses after the Higgs scalar bosons
get vacuum expectation values. For each squark or slepton $\phi$ with
third component of weak isospin $I_{3}^L$ and weak hypercharge
$Y$, this contribution is given by
$$
\Delta_\phi =
M_Z^2 \cos (2 \beta) \left [ \cos^2 \theta_W \, I_{3}^L
- \sin^2 \theta_W \, Y\right ],
\eqno(dsquaredterm)
$$
where
$$
\tan \beta = v_u/v_d
$$
is the ratio between the two expectation values of the Higgs.

Third, there is a supersymmetric contribution which is just equal to the
${\rm (mass)}^2$ of the corresponding quark or lepton. This contribution
is utterly negligible for all but the scalar partners of the top quark.

Finally, there are contributions to the
scalar ${\rm (mass)}^2$ matrix which mix the scalar partners of the
left and right handed squarks and the left and right handed charged sleptons.
These contributions are
again quite negligible for the first two families.

The physical squared masses are obtained from all of the above contributions.
We will use $M_{\tilde q}, M_{\tilde l}$ to denote the physical
masses of squarks and sleptons.
Since  we can neglect Yukawa couplings
the spectrum is arranged in seven distinct groups of degenerate
scalar states $(\sul,\scl)$; $(\sdl,\ssl)$; $(\sur,\scr)$; $(\sdr,\ssr)$;
$(\sel,\smul)$; $(\ser,\smur)$; $(\seln,\smun)$. The
members of each group transform in the same way under
$SU(3)\times SU(2) \times U(1)$.

Now, experimental constraints on
flavor-changing neutral currents are most easily evaded if
the scalar partners of the down and strange squarks are
nearly degenerate, and likewise for the  up and charm squarks and the
selectron and smuon, so there is already indirect experimental evidence
in favor of the hypothesis of a universal mass $m_0$.

With the assumption of a common $m_0^2$, the RG equations for the squarks
and sleptons can be integrated down to experimental scales to yield
$$
\eqalign{
\mmul^2 &= m_0^2 +  C_3 + C_2 + {1\over 36} C_1
+ ({1\over 2} - {2\over 3} \st) M_Z^2 \cos (2 \beta)
\cr
\mmdl^2 &= m_0^2 +  C_3 + C_2  + {1\over 36}  C_1
+ (-{1\over 2} +{1\over 3} \st) M_Z^2 \cos (2 \beta)
\cr
\mmur^2 &= m_0^2 + C_3                  + {4\over 9}  C_1
+ {2\over 3} \st M_Z^2 \cos (2 \beta)
\cr
\mmdr^2 & = m_0^2 +  C_3                 + {1\over 9} C_1
- {1\over 3} \st M_Z^2 \cos (2 \beta)
\cr
\mmel^2 & = m_0^2                + C_2 + {1\over 4}   C_1
+ (-{1\over 2} + \st)  M_Z^2 \cos (2 \beta)
\cr
\mmnl^2 & = m_0^2               +  C_2 + {1\over 4}  C_1
+ {1\over 2} M_Z^2 \cos (2 \beta)
\cr
\mmer^2 & = m_0^2                                  +   C_1
-\st M_Z^2 \cos (2 \beta)
\cr
}
\eqno(sqsl)
$$
where we have added the contributions of the $D^2$-term \(dsquaredterm).
The $C_i$ factors are given by
$$
C_i(t) = \left \lbrace \matrix{ 3/5 \cr 3/4 \cr 4/3 \cr} \right\rbrace
 {1\over 2 \pi^2} \int^{t_X}_{t} {d t}
\, g_i(t)^2 M_i(t)^2
$$
which, after performing the integration can be written as
$$
C_i(t) = \left \lbrace \matrix{ 3/5 \cr 3/4 \cr 4/3 \cr} \right\rbrace
{2 M_{i}^2 (t) \over b_i}\left [ 1 -
 {\alpha_i^2(t_X) \over \alpha_i^2(t)}  \right ].
\eqno(cit)
$$
In \(sqsl), the functions $C_i(t)$ should be evaluated at the
corresponding squark and sleptons mass poles.

Let us suppose for the moment that $\beta$ is known. Then we have
seven physical masses $\mmul$, $\mmdl$, $\mmur$, $\mmdr$, $\mmel$, $\mmnl$,
$\mmer$
which essentially depend on just four
unknown parameters, namely $m_0^2$ and $C_1$, $C_2$, and $C_3$.
Therefore, there should be three independent sum rules which do not
contain the unknown input parameters.

We can immediately use the equations to relate the masses of the squarks and
sleptons which live in the same $SU(2)_L$ doublet:
$$
\eqalignno{
\mmdl^2 -\mmul^2 & = -\cos (2 \beta) M_W^2
&(sra)
\cr
\mmel^2- \mmnl^2 & = -\cos (2 \beta) M_W^2
&(srb)
\cr
}
$$
For the choice $\tan \beta > 1$,
$\cos (2 \beta)$ is negative, so that $\mmdl > \mmul$ and $\mmel > \mmnl$.
Note that these two sum rules do not rely on the assumption of
universal $m_0$ or on the equality of the gaugino masses $M_{i}$
at any initial scale. This is simply because e.g. the left-handed
squarks live in the same irreducible gauge multiplet
before electroweak symmetry breaking. Thus they must have the same $m_0$,
and must be renormalized in the same way down to the electroweak scale.
So the only difference in the masses of $\mmdl$ and $\mmul$ comes
from the electroweak $D$-term, yielding \(sra).
The same argument for the left-handed sleptons yields \(srb).

We also obtain a third sum rule by taking linear combinations of \(sqsl):
$$
2 (\mmur^2 - \mmdr^2 ) + (\mmdr^2 - \mmdl^2 ) +
(\mmel^2 - \mmer^2) = {10\over 3 } \sin^2 \theta_W M_Z^2 \cos 2\beta
\> .
\eqno(src)
$$
This relation {\it does} depend on the assumption of universal
$m_0$, but again does not depend on any particular assumptions
about the gaugino mass parameters. The functions $C_i$ cancel out.
The sum rule \(src) is thus a test of the universality of $m_0$, without
making assumptions about the other input parameters.

The remaining four independent equations can be inverted to yield expressions
for the input parameters in terms of the squark and slepton masses:
$$
\eqalignno{
m_0^2  & = \mmer^2 - 3 (\mmur^2 - \mmdr^2 ) + 4 \sin^2 \theta_W
M_Z^2 \cos 2\beta,
&(mzero) \cr
C_3 & = (\mmdr^2 - \mmer^2 ) + {8\over 3} (\mmur^2 - \mmdr^2 )
- {10\over 3} \sin^2 \theta_W M_Z^2 \cos 2 \beta
&(cthree) \cr
C_2 & = (\mmel^2 - \mmer^2 ) + {9\over 4} (\mmur^2 - \mmdr^2 )
+ \left ( {1\over 2} - {17\over 4} \sin^2 \theta_W \right )
M_Z^2 \cos 2 \beta
&(ctwo) \cr
C_1 & = 3(\mmur^2 - \mmdr^2 )
- 3 \sin^2 \theta_W
M_Z^2 \cos 2 \beta
&(cone) \cr
}
$$
The function $C_3(t)$ varies significantly as a function of scale
even over the range from $M_Z$ to a TeV; here it should be evaluated at a
typical squark mass.

In terms of the gluino mass $\mglu = M_3(t_{\tilde g})$, we have from \(cit)
$$
C_3(t) = {8\over 9} {\mglu^2 \over \alpha_3^2(t_{\tilde g} ) }
\left [ \alpha_3^2 (t) - \alpha_3^2 (t_X) \right ],
\eqno(cthreeal)
$$
where we have used \(mscaling). With the assumption of a GUT, one can
further require that all three gaugino masses be the same at $M_X$. While
there are theories, derived from superstrings, where the unification
of the gauge couplings does not imply the equality of the gaugino masses at
that scale, in the following we may choose to assume the following GUT relation
$$
M_{i} (t_X) \equiv m_{1/2} .
\eqno(gauginoassumption)
$$
It follows from \(mscaling) that
$$
M_i(t) = \alpha_i(t) {\mglu \over \alpha_3 (t_{\tilde g}) } \>.
$$
Then the seven equations for the squarks and sleptons of a light family are
expressed in terms of two mass parameters $m_0$ and  $\mglu$,
and the angle $\beta$. We can now test the assumption of equal gaugino
masses at $t_X$, since it implies that
$$
C_1(t) = {2\over 11} {\mglu^2 \over \alpha_3^2(t_{\tilde g} ) }
\left [ \alpha_1^2 (t_X) - \alpha_1^2 (t) \right ],
\eqno(ctwoal)
$$
$$
C_2(t) = {3\over 2} {\mglu^2 \over \alpha_3^2(t_{\tilde g} )   }
\left [ \alpha_2^2 (t_X) - \alpha_2^2 (t) \right ].
\eqno(coneal)
$$
We can estimate the values for $C_1$, $C_2$ and $C_3$ in terms of
the gluino mass, by assuming that at the unification scale,
$\alpha_1(t_X) = \alpha_2(t_X) = \alpha_3(t_X) = .04$ .
These estimates depend strongly on the value of the QCD coupling constant
at low energies,
which we take to be $ .08 < \alpha_3(t_{\tilde g}) < .11 $,
with the lower (upper) bound corresponding to a heavy (light) gluino.
[We take $\alpha_3(M_Z) = .115$ .]
Then for $C_1$ and $C_2$ (in the hundreds of GeV range) we estimate
$$
.020 \> \mglu^2 < C_1 < .037 \> \mglu^2 ,
\eqno(coneest)
$$
$$
.058 \> \mglu^2 < C_2 < .115 \> \mglu^2  .
\eqno(ctwoest)
$$
The other parameter $C_3$ should be evaluated at a typical squark mass
scale $t_{\tilde q}$. If $m_0$ is small, then $t_{\tilde q}$ is slightly
less than $t_{\tilde g}$, and we find
$$
.67\> \mglu^2 < C_3 < .80 \> \mglu^2 \> .
\eqno(cthreeest)
$$
For larger values of $m_0$, the squarks will be heavier than the
gluino, so from \(cthreeal) we find that the estimate for $C_3$ decreases.
However, a reasonable general range is
$$
.35\> \mglu^2 < C_3 < .80 \> \mglu^2  \> .
\eqno(cthreeestlib)
$$
More precision in the expected values of the functions $C_i$
must await a better determination of the gauge couplings as a function of
scale in the sparticle mass range. The discovery of the
sparticles will then allow the RG thresholds to be implemented,
and the idea of gaugino mass unification tested.

{}From equations \(cthree) and \(ctwo), we obtain
$$
\eqalign{
C_2 - {3\over 4} C_1 & = \mmel^2 - \mmer^2 +
({1\over 2} - 2 \sin^2 \theta_W ) M_Z^2 \cos (2\beta )
\cr
C_3 - {8\over 9} C_1 & = \mmdr^2 - \mmer^2
- {2 \over 3} \sin^2 \theta_W M_Z^2 \cos (2\beta )
\> .
}
$$
By taking their ratio we arrive at a sum rule which is independent
of the gluino mass, namely (in GeV$^2$)
$$
\mmel^2 - \mmer^2 =
{C_2 - {3\over 4 } C_1 \over C_3 - {8\over 9} C_1 }
\left [ \mmdr^2 - \mmer^2  - (36)^2 \cos (2\beta) \right ]
+ (20)^2 \cos (2\beta)
\eqno(srd)
$$
or
$$
\mmel^2 - \mmer^2 =
[.07\> {\rm to} \> .31 ]
( \mmdr^2 - \mmer^2  )
+ [ (18)^2 \> {\rm to} \> (0)^2 ] \cos (2\beta)
\> .
\eqno(srdd)
$$
which tests the unification hypothesis for the gauge couplings
and gaugino masses at $M_X$.
The present large uncertainty in the numerical value of \(srdd)
is due partly to the uncertainty in $\alpha_3 (M_Z)$ but more importantly
to our lack of knowledge of the sparticle masses, which we need
to tell us where to evaluate $C_{1,2,3}$ and where the thresholds are.
At any rate, this formula shows that $\mmdr > \mmel$.

\subhead{4. THIRD FAMILY SQUARKS AND SLEPTONS}
\taghead{4.}

Sum rules involving the masses of third family squarks and sleptons are more
complicated because of the Yukawa couplings.
The presence of $y_t^2 {\Sigma }_t^2$,
$y_b^2 {\Sigma }_b^2$ and $y_\tau^2 {\Sigma }_\tau^2$ in the RG equations
\(rgestop) and \(rgestau), and the complicated RG evolution of the Higgs mass,
makes these equations hard to integrate in a useful form. In addition, there
are mixing terms in the mass matrices for the third family squarks and
sleptons.

The mass-squared
matrix for the top squarks is given by
$$
\pmatrix{m^2_{\tilde t_L}+m^2_t+\Delta_{\tilde t_L} &
m_t (A_t  + \mu \cot \beta)\cr
m_t (A_t  + \mu \cot \beta)&
m^2_{\tilde t_R} + m^2_t + \Delta_{\tilde t_R}\cr}
\eqno(topmatrix)
$$
and that of the bottom squarks by
$$
\pmatrix{m^2_{\tilde b_L}+m_b^2+\Delta_{\tilde b_L} &
m_b (A_b  + \mu \tan \beta)\cr
m_b (A_b  + \mu \tan \beta)&
m^2_{\tilde b_R} +m_b^2 + \Delta_{\tilde b_R}\cr}
\> .
\eqno(bottommatrix)
$$
Despite these complications, with further
assumptions concerning the relative magnitudes of the Yukawa couplings of the
third family we can deduce some new sum rules for the third family squark
masses.

The validity of the radiative electroweak symmetry breaking with a top quark
mass much larger than the bottom quark mass puts restrictions on the relative
magnitudes of $y_b$ and $y_t$. For $y_t \ll y_b$, the radiative breaking
scenario implies $m_b > m_t$. For $y_t \sim y_b$, the two Higgs develop
similar vacuum expectation values, which in turn implies $m_b \sim m_t$ in the
absence of fine-tuning.
This leaves us with only one viable possibility, $y_t \gg y_b$. In this case,
the radiative mechanism naturally favors a larger vacuum expectation value
for $H_u$ which couples to the top, yielding a consistent picture when
$\tan \beta \ll m_t/m_b$. It is amusing to note that in
$SO(10)$, this hierarchy of Yukawa couplings
has a natural explanation provided that
a $\bf 126$ Higgs couples more strongly than the $\bf 10$ to the top.

In the following, we therefore neglect $y_b$.
(Numerical work[\cite{sla}] indicates that this is a reasonable
approximation in realistic models for
$\tan \beta$ less than about 10.)
Then there is no mixing in the bottom squark mass matrix, and
$\tilde b_L$ and $\tilde b_R$ are still the mass eigenstates.
Thus, $\tilde b_R$ is degenerate with $\tilde d_R$ to a good approximation.
In addition, from the RG equation  for the running masses
$$
16 \pi^2 {d\over dt} (\mbl^2 - \mdl^2 ) = 2 y_t^2 {\Sigma}_t^2,
\eqno(bdsplit)
$$
which has a positive RHS, we note the inequality
$$
\mmbl < \mmdl
\eqno(blineq)
$$
for the physical masses.
It is also true that ${\tilde d}_R$ is lighter than $\tilde d_L$,
but the relative placement of $\tilde b_L$ and $\tilde d_R$ cannot
be determined without more detailed knowledge of the input parameters.

In the stop sector, the analysis is different, because
$m^2_{\tilde t_L}$ and $m^2_{\tilde t_R}$, which
are the result of running
the universal value $m_0^2$ from $\Lambda$ down to the electroweak scale,
are not the actual mass eigenvalues.
The mass eigenstates $\tilde t_1, \tilde t_2$
of the top squark system are found by diagonalizing
the matrix \(topmatrix), whose eigenvalues are the
physical squared masses $M^2_{\tilde t_1}, M^2_{\tilde t_2}$.
The sum of the (mass)$^2$
eigenvalues is just the sum of the diagonal entries in \(topmatrix).
So, by taking the trace, we find
$$
M^2_{\tilde t_1} +
M^2_{\tilde t_2} = m_{{\tilde t}_L}^2 +
m_{{\tilde t}_R}^2 + 2 m_t^2 + {1\over 2} M_Z^2 \cos
2 \beta \> .
$$
We observe that there are two linear combinations
of $m^2_{\tilde t_L}$, $m^2_{\tilde t_R}$, and  $m^2_{\tilde b_L}$
for which the terms involving $y_t^2 {\Sigma}_t^2$ in the RG equation
\(rgestop)
cancel out, and which therefore evolve like their
counterparts from the first two families.
One is the linear
combination $m_{\tilde t_L}^2 + m_{\tilde t_R}^2 - 3 m_{\tilde b_L}^2 $
which runs exactly as the combination
$m_{\tilde u_L}^2 + m_{\tilde u_R}^2 - 3 m_{\tilde d_L}^2 $ from
the first two families.
The $D$-term contributions  to these two combinations
are equal. We therefore obtain the interesting new sum rule
$$
\mmtone^2 + \mmttwo^2 - 3 \mmbl^2 - 2 m_t^2 =
            \mmul^2 + \mmur^2 - 3 \mmdl^2
\eqno(sre)
$$
which relates masses of the squarks and quarks of the third family
with the masses of the squarks of the first two families, without
involving any input parameters.

If the top squark matrix is diagonalized by a rotation through
an angle $\varphi $, we have
$$
( M_{\tilde t_1}^2 - M_{\tilde t_2}^2    )
\sin (2 \varphi )
= 2 m_t (A_t + \mu \cot \beta ),
$$
$$
( M_{\tilde t_1}^2 - M_{\tilde t_2}^2    )
\cos (2 \varphi )
= m_{\tilde t_L}^2 - m_{\tilde t_R}^2 + ({1\over 2} - {4\over 3} \sin^2
\theta_W ) M_Z^2 \cos 2 \beta
\> .
$$
Eliminating the angle $\varphi $, we obtain
$$
( M_{\tilde t_1}^2 - M_{\tilde t_2}^2    )^2
= 4 m_t^2 (A_t +\mu \cot \beta )^2 +
\left [ m_{\tilde t_L}^2 - m_{\tilde t_R}^2
+ ({1\over 2} - {4\over 3} \sin^2 \theta_W ) M_Z^2 \cos 2\beta
  \right ]^2
\>.
$$
Noting that the combination $m_{\tilde t_L}^2 - m_{\tilde t_R}^2 $
runs across the scales exactly like
$-\mbl^2 + \mul^2 + \mdl^2 - \mur^2$, and taking into account
the $D$-term contribution leads to our second sum rule
for the third-family squarks:
$$
( M_{\tilde t_1}^2 - M_{\tilde t_2}^2    )^2
= 4 m_t^2 (A_t +\mu \cot \beta )^2 +
\left [ \mmbl^2 - \mmdl^2 - \mmul^2 + \mmur^2 \right ]^2
\eqno(srf)
$$
This equation provides a lower bound on the splitting between
the top-squark masses, and illustrates how the parameters $A_t $ and
$\mu$ contribute to the splitting in the stop sector. We can also express
the top-squark mixing angle in terms of the physical masses by
$$
\cos 2 \varphi \> = \>
{  \mmdl^2 - \mmbl^2 + \mmul^2  - \mmur^2  \over
M_{\tilde t_1}^2 - M_{\tilde t_2}^2 } \> .
\eqno(costwophi)
$$
If the stop mixing angle $\varphi$ can also be measured by other means,
this may provide another interesting test.

{}From the form of the mixing matrix of the bottom squarks, it may be that
neglecting $y_b$ is inappropriate if $\mu$ and/or
$\tan \beta$ is very large. In that case,
$\tilde b_R$ and $\tilde d_R$ are no longer degenerate, and our sum rules
may have to be modified.

By the same token, for large $\tan \beta$ and $\mu$, one should also take into
account the left-right mixing and the effect of the tau Yukawa
coupling in the third family slepton sector.
The stau (mass)$^2$ matrix is given by
$$
\pmatrix{m^2_{\tilde \tau_L}+m^2_\tau+\Delta_{\tilde \tau_L} &
m_\tau (A_\tau  + \mu \tan \beta)\cr
m_\tau (A_\tau  + \mu \tan \beta)&
m^2_{\tilde \tau_R} + m^2_\tau + \Delta_{\tilde \tau_R}\cr}
\> .
\eqno(taumatrix)
$$
For very large $\mu$ and $\tan \beta$, the splitting between $\tilde\tau_L$ and
$\tilde\tau_R$ will be increased somewhat by the left-right mixing terms.
Since the mixing angle is always small,
we use the same names $\tilde \tau_L$ and $\tilde \tau_R$ for the mass
eigenstates as for the gauge eigenstates.
Also, $m^2_{\tilde \nu_\tau}$, $m^2_{\tilde \tau_L}$ and
$m^2_{\tilde \tau_R}$ are pushed lower because of the terms proportional to
$y_\tau^2 \Sigma_\tau^2$ in \(rgestau).
Since
$$
16 \pi^2 {d\over dt}( m^2_{\tilde \nu_\tau} - m^2_{\tilde \nu_e} )
= 2 y_\tau^2 \Sigma_\tau^2 > 0
$$
we know that
$$
M_{\tilde \nu_\tau} < M_{\tilde \nu_e} \> .
\eqno(sntauineq)
$$
By taking the traces of the stau (mass)$^2$ matrix and its selectron
counterpart, and noting that the renormalization of the combination
$m_{\tilde \tau_L}^2 + m_{\tilde \tau_R}^2 - 3 m_{\tilde \nu_\tau}^2 $ does
not contain the $\tau$ Yukawa coupling, we derive the sum rule
for the physical masses:
$$
M^2_{\tilde \tau_L}+
M^2_{\tilde \tau_R}
-3 M^2_{\tilde \nu_\tau}=
\mmel^2+\mmer^2-3 M^2_{\tilde \nu_e}
\> .
\eqno(stausq)
$$
We see from \(sntauineq) and \(stausq) that the center of mass-squared
for the staus is less than that of the selectrons.
Numerical work[\cite{sla}] shows that typically
$\tilde \tau_L$ is slightly heavier than
$\sel$ and $\tilde \tau_R$ is lighter than $\ser$ and $\tilde \nu_\tau$
is lighter than $\tilde \nu_e$ (by at most a few GeV in each
case) when both $\tan \beta$ and $\mu$ are large.

\subhead{5. CHARGINOS}
\taghead{5.}

The chargino sector consists of the fermionic partners of the
charged electroweak gauge bosons and of the charged Higgs scalar bosons.
The mass matrix is[\cite{char}]
$$
( {\tilde W}^+ \> {\tilde H}^+ )
\pmatrix{
M_2 & {\sqrt 2} M_W \cos \beta \cr
{\sqrt 2} M_W \sin \beta & \mu \cr }
\pmatrix{ {\tilde W}^- \cr -{\tilde H}^- }
+ {\rm c.c. }
$$
This mass matrix describes two charged Dirac fermion mass eigenstates
${\tilde C}_1$ and ${\tilde C}_2$ with masses
$$
M^2_{{\tilde C}_{1,2}}
=
{1\over 2} \left [ (M_2^2 + \mu^2 + 2 M_W^2) \pm
\sqrt{(M_2^2 + \mu^2 + 2 M_W^2 )^2 - 4 ( \mu M_2 - M_W^2 \sin 2 \beta)^2 }
\right ] .
$$

If the gluino mass is known, then the gaugino mass parameter
$M_2$ is $  \alpha_2 (M_{\tilde g} / \alpha_3)$, with $\alpha_3$
taken at the gluino mass scale and $\alpha_2$ evaluated
self-consistently at $M_2$. Thus,
measurement of the two chargino masses in principle determines the
two unknown parameters $\mu$ and $\beta$.
In fact, the sum of the squares of the charginos depends only on
$\mu$ and not on $\beta$:
$$
M^2_{{\tilde C}_1} + M^2_{{\tilde C}_2} = M_2^2 + 2 M_W^2 + \mu^2 \> .
\eqno(charsqsum)
$$
Also the product of the chargino eigenstates is given simply by
$$
M_{{\tilde C}_1} M_{{\tilde C}_2} = \mu M_2  - M_W^2 \sin 2\beta \> .
\eqno(charprod)
$$
Using these equations, and a measurement of the physical masses
of ${\tilde g}$, ${{\tilde C}_1}$, ${{\tilde C}_2} $, and couplings
$\alpha_2$, $\alpha_3$, one can solve for $\mu$ from \(charsqsum)
and then for $\sin 2 \beta$ from \(charprod).
In a region $\tan \beta \gg 1$, this provides a more sensitive
measure of the angle $\beta$ than can be obtained in the squark and
slepton sector via eqs.~\(sra) or \(srb). The value of $\beta$ determined
by the chargino sector masses from \(charsqsum) and \(charprod) should
therefore be used as an input for the squark and slepton sum rules.

In realistic models, it often happens that the chargino masses are
close to being degenerate with two of the four neutralino masses.
As we will see, this can be explained by considering the limit in which
$M_Z$ is small compared to $\mu \pm M_2$, so that electroweak
symmetry breaking can be treated as a perturbation in the chargino
and neutralino mass matrices. From this point of view,
the masses of the charginos are given to the lowest non-trivial order by
$$
\eqalign{
M_{{\tilde C}_1} &= M_2 -
{ M_W^2 (M_2 + \mu \sin 2 \beta ) \over \mu^2 - M_2^2 }
\cr
M_{{\tilde C}_2}
& = \mu + {M_W^2 (\mu + M_2 \sin 2 \beta) \over \mu^2 - M^2_2 }
\> .
}
\eqno(pertchar)
$$
The eigenstate ${\tilde C}_1$ is mostly wino and the eigenstate
${\tilde C}_2$ is mostly charged Higgsino in this limit.

\subhead{6. NEUTRALINOS}
\taghead{6.}

The neutralino sector consists of the fermionic partners of the neutral
electroweak gauge bosons and of the neutral Higgs scalar bosons. Electroweak
symmetry breaking introduces mixing between these states. The mass
spectrum and mixing angles are determined by the mass matrix
$$
\pmatrix{
M_1 & 0 & -M_Z \cos \beta \sin \theta_W & M_Z \sin \beta \sin \theta_W \cr
0 & M_2 & M_Z \cos \beta \cos \theta_W & -M_Z \sin \beta \cos \theta_W \cr
-M_Z \cos \beta \sin \theta_W & M_Z \cos \beta \cos \theta_W & 0 & -\mu \cr
M_Z \sin \beta \sin \theta_W & -M_Z \sin \beta \cos \theta_W & -\mu & 0
\cr
}
\eqno(neutmass)
$$
in the basis $({\tilde B},{\tilde W}^0, -i {\tilde H}^0_u,-i {\tilde H}^0_d )$.
The neutralino mass eigenvalues thus satisfy the characteristic equation
$$
\eqalign{ 0= &
\lambda^4 - \lambda^3 (M_1+M_2) + \lambda^2 (M_1 M_2- \mu^2 - M_Z^2 )
\cr & + \lambda ( \mu^2 [M_1+M_2]  +  M_W^2 [M_1 +  M_2 \tan^2 \theta_W ]
- \mu M_Z^2 \sin 2 \beta )
\cr & - \mu^2 M_1 M_2 +  \mu M_W^2 [M_1 + M_2 \tan^2 \theta_W ] \sin 2 \beta
\>.
}
\eqno(neutchareq)
$$
The exact analytical expressions for the mass eigenvalues are quite
complicated and not very illuminating. However, we can still make some
relatively simple statements about the spectrum of neutralinos in the form
of sum rules for the physical masses.

A simple relation governs the product of the neutralino masses,
which is equal to the determinant of \(neutmass), and from
\(neutchareq) is given by
$$
M_{{\tilde N}_1} M_{{\tilde N}_2} M_{{\tilde N}_3} M_{{\tilde N}_4} =
- \mu^2 M_1 M_2  + \mu M_W^2 [M_1 +  M_2 \tan^2 \theta_W ] \sin 2 \beta
\> .
\eqno(neutdet)
$$
This will provide an independent test of the values of $\mu$ and $\beta$
obtained from the chargino spectrum via \(charsqsum) and \(charprod).

Knowledge of the sign of the determinant of the neutralino mass matrix
is important in the derivation of neutralino mass sum rules.
For $\mu<0$, the determinant is obviously negative, and it is easy to show
that one of its eigenvalues is negative and the other three positive.
If $\mu >0$, the determinant is still negative as long as
$ \mu M_2 > 1.6 M_W^2 \sin 2 \beta $
where we have used the fact that $M_1$ is approximately $.5 M_2$.
However, the present experimental bounds on the chargino masses
($M_{\tilde C_i} \ge M_Z/2$) and on the gluino mass ($\mglu > 100$ GeV)
still allow for the existence of a very
restricted range of parameters for which the determinant is positive,
namely $$.45 < \tan \beta < 2.2\> ,$$
$$M_2^2 + \mu^2 < 1.5 M_W^2\> ,$$
$$ \mu M_2 < .69 M_W^2\> .$$
Note that LEPII can rule out the existence of this very small window by
failing to detect any chargino lighter than the $W$. Also, the window shrinks
rapidly as the lower limit on the gluino mass increases, disappearing
entirely for $\mglu$ greater than about $300$ GeV.

The sum of the eigenvalues of the neutralino mass matrix is equal to its trace,
which is $M_1+M_2$, and thus does not depend on $\mu$ or $\beta$.
In most of the allowed parameter space, where the determinant is negative,
exactly one of the eigenvalues is negative.
We call the neutralino eigenstate of \(neutmass) which corresponds to the
negative eigenvalue the ``flipped" neutralino. Then by relating
$M_1$ and $M_2$ to the gluino mass, we arrive at the simple sum rule
$$
|M_{{\tilde N}_1}| + |M_{{\tilde N}_2}| + |M_{{\tilde N}_3}|-|M_{{\tilde N}_4}|
\> = \>  (\alpha_1 + \alpha_2 ) {M_{\tilde g}  \over  \alpha_3 }
\eqno(srg)
$$
where ${\tilde N}_4$ is the flipped neutralino.
In this expression, $\alpha_3$ should be
evaluated at the gluino mass scale, while $\alpha_1$ and $\alpha_2$
should be evaluated at the neutralino mass scale. Typically,
one then finds very roughly that $(\alpha_1 + \alpha_2)/\alpha_3 \approx
.5$ in eq.~\(srg).  We suggest that in future numerical work on
the sparticle spectrum, it would be useful to specify not only the
masses of the four neutralinos, but also which of them is the
flipped neutralino in the sense discussed here.

In the very unlikely case discussed above of a positive determinant for
\(neutmass), the term proportional to $\lambda^2$ in the characteristic
equation, $M_1 M_2 - \mu^2 - M_Z^2$ is negative, which implies that two of the
eigenvalues are negative and two are positive.
Then the trace sum rule \(srg) would be replaced by
$$
|M_{{\tilde N}_1}| + |M_{{\tilde N}_2}|  -|M_{{\tilde N}_3}|-|M_{{\tilde N}_4}|
\> = \> (\alpha_1 + \alpha_2 ) {M_{\tilde g}  \over  \alpha_3 } \> .
\eqno(sri)
$$

The sum of the squares of the neutralino masses is given
by the trace of the square of \(neutmass):
$$
M^2_{{\tilde N}_1} + M^2_{{\tilde N}_2} +
M^2_{{\tilde N}_3} + M^2_{{\tilde N}_4} =
M_1^2 + M_2^2 + 2 \mu^2 + 2 M_Z^2 \> .
\eqno(neutsqsum)
$$
Combining this with the chargino (mass)$^2$ relation \(charsqsum),
and writing $M_1$ and $M_2$ in terms of the gluino mass,
we arrive at the sum rule
$$
2 ( M^2_{{\tilde C}_1} + M^2_{{\tilde C}_2} )
- ( M^2_{{\tilde N}_1} +
M^2_{{\tilde N}_2} + M^2_{{\tilde N}_3} + M^2_{{\tilde N}_4} ) =
(\alpha_2^2 - \alpha_1^2){ M_{\tilde g}^2 \over \alpha_3^2}  + 4 M_W^2
- 2 M_Z^2 \> .
\eqno(srh)
$$
In this formula $\alpha_3$ should again be evaluated at the gluino mass scale.
A corollary of \(srh) is that the average squared mass of the
neutralinos is always less than the average squared mass of the charginos.
The virtue of \(srg) and
\(srh) is that all dependence on input parameters
has been eliminated in favor of  physical masses
and coupling constants.
They should hold in general as long as the GUT assumption
relating the gaugino mass parameters $M_1$, $M_2$ and $M_3$ is true,
notwithstanding the complicated dependence of the neutralino and chargino
mixings on the unknown parameters $\mu$ and $\beta$.

The mass scale of the neutralino
sector is set by $\mu$, $M_1$, and $M_2$. In fact, with $M_Z=0$,
the neutralino mass eigenvalues of \(neutmass)
are $M_1$, $M_2$, $\mu$ and $-\mu$, and there is no mixing
between gauginos and Higgsinos. Now suppose that we turn
on electroweak symmetry breaking. Then, expanding in $M_Z$,
the neutralino mass eigenvalues are perturbed to
$$
\eqalign{
M_{{\tilde N}_1} & = M_1 -
{ M_Z^2 \sin^2 \theta_W (M_1 + \mu \sin 2 \beta ) \over \mu^2 - M_1^2 }
\cr
M_{{\tilde N}_2} & = M_2 -
{ M_W^2 (M_2 + \mu \sin 2 \beta ) \over \mu^2 - M_2^2 }
\cr
M_{{\tilde N}_3} & = \mu  +
{ M_Z^2  (1+ \sin 2 \beta) (\mu - M_1 \cos^2 \theta_W -M_2 \sin^2 \theta_W )
\over 2 (\mu-M_1) (\mu - M_2) }
\cr
M_{{\tilde N}_4} & = -\mu  -
{ M_Z^2  (1- \sin 2 \beta) (\mu + M_1 \cos^2 \theta_W +M_2 \sin^2 \theta_W )
\over 2 (\mu + M_1) (\mu + M_2) }
\> . \cr
}
\eqno(pertneut)
$$
These expressions generalize the ones given in [\cite{AN2}].
They are valid so long as $M_Z$ is small compared to $\mu \pm M_{1,2}$.
(In cases like
$\mu \approx \pm M_2 > M_Z$  the above expressions are not reliable,
but one can do almost-degenerate perturbation theory to find the
neutralino mass eigenvalues.)
If we also assume that $|\mu|$ is larger than $M_{1,2}$, then the
LSP is ${\tilde N}_1$, since $M_1$ is typically about half of $M_2$.
The physical neutralino masses are the absolute values of these quantities.
In \(pertneut), the flipped neutralino is ${\tilde N}_4$ if
$\mu$ is positive and  is ${\tilde N}_3$ if $\mu$ is negative.
The electroweak interactions split the degeneracy between the neutralinos
${\tilde N}_3$ and ${\tilde N}_4$.
By comparing \(pertneut) with \(pertchar), we see that the chargino
${\tilde C}_1$ and the neutralino ${\tilde N}_2$ are exactly degenerate
to this order in the expansion in $M_Z^2$:
$$
M_{\tilde C_1} = M_{\tilde N_2} +
{\cal O}\left ( {M_Z^2\over \mu^2 - M_{1,2}^2} \right )^2 \> \> .
\eqno(cndega)
$$
Also, the neutralino
${\tilde N}_3$ is often quite close in mass to the other chargino
${\tilde C}_2$; they are exactly degenerate in the limit of no
electroweak breaking and the corrections from this limit turn
out to be similar. For example, in the large $\mu$ limit, one has
$$
M_{\tilde C_2} - M_{\tilde N_3} =
{M_Z^2 \over \mu} \left [ \cos^2 \theta_W - {1+ \sin 2 \beta \over 2}
\right ] \> \> .
\eqno(cndegb)
$$
For $\tan \beta = $  a few, this happens to be numerically  small.
Numerical calculations have shown that these coincidences
are quite good, even when the expansion in $M_Z^2$ is not so reliable.

\subhead{7. DISCUSSION}
\taghead{7.}

Supersymmetry  predicts[\cite{lighthiggs}]
the existence of a light Higgs scalar, which should be discovered at
LEPII if its mass is less than about 90 GeV (perhaps 118 GeV),
and at the SSC or LHC
otherwise. However, discovery of a light Higgs by itself will
neither confirm nor deny the existence of supersymmetry, since it
can also be a feature of non-supersymmetric models. The first definitive
experimental signal of supersymmetry may very well turn out to be the
discovery of the gluino at a hadron collider. Because the gluino
is a color octet, it should be copiously produced at the SSC and LHC, and
its mass measured.

In the following, we adopt the GUT assumption for the gaugino mass
parameters. With the gluino mass known, this fixes
the gaugino mass parameters $M_2$ and $M_1$ which appear in the
chargino and neutralino sector, {\it and } the functions $C_1,C_2,C_3$
appearing in the formulas for the squark and slepton masses.
Numerically, one typically has $M_1 \approx .17 \mglu$, $M_2 \approx
.33\> \mglu$,
and the ranges for $C_1$, $C_2$, $C_3$ are given by
\(coneest)-\(cthreeestlib).

\noindent{\bf Squarks and Sleptons  }

Knowledge of the gluino mass  determines
the splittings in the squark and slepton (mass)$^2$
spectrum. The overall scale  in this spectrum is set by the
universal parameter $m_0^2$, which does not appear in the splittings
of the squared masses. The spinless sparticles of the first two
families generally arrange themselves into three ``bands":

\noindent $\bullet$ The lightest of these bands contains the three right-handed
sleptons,
$(\tilde e_R, \tilde \mu_R, \tilde \tau_R)$. This is a consequence of
$C_1 < C_2, C_3$. The mass scale for this band of right-handed sleptons
is set by $m_0$ and $C_1$. For larger values of $\tan \beta$,
$\tilde \tau_R$ is slightly lighter than $(\tilde e_R, \tilde \mu_R)$.

\noindent $\bullet$ The middle band contains the three degenerate
left-handed charged sleptons,
$(\tilde e_L, \tilde \mu_L, \tilde \tau_L)$ and the three sneutrinos
$(\tilde \nu_{e }, \tilde \nu_{\mu }, \tilde \nu_{\tau })$,
with a slightly different mass determined by the sum rule  \(srb).
This splitting within the band is most pronounced if both $m_0$ and
the gluino mass are in the lower part of their allowed ranges, because
then the $D$-term contribution is relatively more significant compared
to $m_0^2$ and $C_2$.
The splitting between the light band of right-handed sleptons and
the middle band of left-handed sleptons is governed by the value
of $C_2$ via \(ctwo). The splitting between the two lower bands
is more significant if
the gluino mass is relatively large compared to $m_0$, as in
``no-scale" models.
For large $\tan \beta$, $\tilde\nu_\tau$ is lighter than $\tilde \nu_e$
and $\tilde \tau_L$ is slightly heavier than $\tilde e_L$.

\noindent $\bullet$
The heaviest band contains all of the squarks of the first two families
(and $\tilde b_R$ if $\tan\beta$ is not too large).
The essential reason they are all heavier than the sleptons,
and why they congregate in a band, is because they all obtain a large common
contribution from the RG equation which is $C_3 \gg C_1, C_2, M_Z^2$.
Within this band, there is a small splitting between the groups
$(\tilde u_L , \tilde c_L )$ and $(\tilde d_L , \tilde s_L ) $
as mandated by the sum rule \(sra). The splitting (within the band) between
$(\tilde u_R , \tilde c_R )$ and $(\tilde d_R , \tilde s_R ) $
is small,
giving a measure of the value of $C_1$ after the $D$-term contribution
in \(cone) is taken into account. This can be interpreted in terms of the
custodial symmetry of the standard model. For
$\tan \beta > 1$, the $D$-term contribution
to the splitting between right-handed up and down-type squarks happens
to have the opposite sign from the RG contribution from $C_1$, increasing their
tendency to be degenerate in mass. Numerically one has
(in GeV$^2$)
$$
\mmur^2 - \mmdr^2 \approx  (.1 \> \mglu )^2 + (43)^2 \cos (2\beta)
\> .
$$
Knowing the value of $C_3$
tells us the approximate splitting of the heaviest band of squarks
from the lighter bands of left-handed sleptons and of right-handed sleptons
through \(srd).

These qualitative features of the spectrum of the squarks and sleptons
of the first two families change drastically depending on the relative
values of the gluino mass and the input parameter $m_0$.

In the ``no-scale" limit $m_0 \ll \mglu$, the three bands should be well
separated in mass, with a discernable structure within each band. In this
case, the sum rules \(sra), \(srb), and \(src), which do not rely on the
input parameters, can be tested. In addition, the measurement of the separation
between the bands directly tests the hypothesis of equal input
gaugino masses.

In the opposite ``anti-no-scale" limit, $m_0 \gg \mglu$, $m_0$ dominates
the mass spectrum, all the bands are bunched together, and any hint of
the structure within the bands disappears. The most
extreme versions  of this limit  are already ruled out, because of lower
limits on the mass of the gluino.

The squarks of the third family are not degenerate with those of the first
two families, because the Yukawa couplings are significant.

The values of the stop masses $\tilde t_1$ and $\tilde t_2$ are the result of
several competing effects. For one, the term proportional
to $y_t^2 \Sigma_t^2$ in the RG equations pushes the masses
lower compared to their counterparts from the first two families.
There is  also a positive contribution for the top squarks of
magnitude $m_t^2$. Finally, the left-right cross-terms for the top squarks
introduces a mixing depending on $A_t + \mu \cot\beta$,
which increases one eigenvalue and lowers the other.

The bottom squark mass eigenstates are also different from their
counterparts $\sdl$ and $\sdr$ because of three effects. First, $\mbl^2$
is smaller than $\mdl^2$ because of the term proportional to $y_t^2 \Sigma_t^2$
in the RG equations. Second, the terms proportional to $y_b^2 \Sigma_b^2 $
in the RG equations push both $\mbl^2$ and $\mbr^2$ lower than $\mdl^2$
and $\mdr^2$.
Finally, the left-right cross term introduces a mixing of $\sbl$ and $\sbr$
depending on $m_b(A_b + \mu \tan \beta)$, so that the splitting between the
true bottom squark
mass eigenstates is larger than the splitting between $\sdl$ and $\sdr$.
The latter two effects are only significant if $\tan \beta $ is comparable to
$m_t/m_b$, which we have noted is difficult to reconcile with
the radiative electroweak breaking mechanism. In the usual case where
$\tan \beta$ is at most about 10, $\sbr$ is degenerate with $\sdr$, and
the bottom squark mixing is negligible.

The two sum rules \(sre) and \(srf)
allow us to analyze the qualitative features of the spectrum.
When the first two family squarks are clumped together, we can rewrite
\(sre) in the form
$$
M_{\tilde t_1}^2 + M_{\tilde t_2}^2 = 2 M_{\tilde b_L}^2   + 2 m_t^2 +
 (M^2_{\tilde b_L} - M^2_{\tilde q})
$$
where $\tilde q$ is a generic squark from the first two families.
We see that the location of the center of mass squared
of $\tilde t_1$ and $\tilde t_2$ is determined by the amount by which
$\tilde b_L$ is lower than the main squark band.
Similarly, the mass squared difference sum rule \(sre) becomes effectively
$$
\left ( M_{\tilde t_1}^2 - M_{\tilde t_2}^2\right )^2 =
4 m_t^2 (A_t + \mu \cot \beta )^2 + \left (
M_{\tilde q}^2 - M_{\tilde b_L}^2  \right )^2
$$
indicating a lower bound for the splitting between $\tilde t_1$ and
$\tilde t_2$ which is determined by that between $\tilde b_L$ and
the main squark band. Thus for a small difference between
$\tilde  b_L$ and the main squark band, the split between $\tilde t_1$
and $\tilde t_2$ may be small, if $A+\mu\cot\beta$ is small. However,
for large values of $A$ or $\mu$, the difference may be substantial.
One then expects $\tilde t_2$ to be above the main squark band, and
$\tilde t_1$, $\tilde b_L$ below. In the ``anti-no-scale" limit,
$\tilde b_L$'s mass can be much lower than the main squark band.
In this case, the center of mass squared of $\tilde t_1$ and $\tilde t_2$
is lower than $\tilde b_L$ which is itself much lower than the rest
of the squarks. Also, the split between $\tilde t_1$ and $\tilde t_2$
may be very large, depending on the crossing term.
If it is large enough, $\tilde t_2$ will be heavier than $\tilde b_L$.

\noindent{\bf Charginos and Neutralinos}

The masses of the charginos and neutralinos are highly correlated with
each other, and are primarily determined by the input
parameters $\mu$ and $\sin 2 \beta$, as well as by the gluino mass.
In the limit when $M_W^2 \ll \mu^2 - (.1\>  \mglu^2)$, one of the charginos
is degenerate with a neutralino, from eq.~\(cndega).
The other chargino is also usually
close in mass to another neutralino, especially if $\tan \beta$ is in a range
near 3 or 4, as we see from eq.~\(cndegb).

The lightest of the neutralinos (LSP) is absolutely stable. In order to
avoid cosmological problems, $\mu$ and $\mglu$ cannot both be arbitrarily
large. The center of masses of the neutralino is smallest when the
flipped neutralino is the LSP, as we see from the trace sum rule
\(srg). If $\mu$ is large compared to $M_1 \approx .17\> \mglu$ and
$M_2 \approx .33 \> \mglu$, then $\tilde N_1$ in \(pertneut) is the LSP,
and the trace sum rule still tells us about the spread of the neutralino
masses, and tests the idea of gaugino mass unification. The sum rule \(srh)
indicates that the center of mass squared of the charginos
is higher than that of the neutralinos.

We have mentioned in Section 4 that $\mu$ and $\sin 2 \beta$ are likely to
be measured by the chargino masses [see eqs.~\(charsqsum)
and \(charprod)]. Knowing $\mu$ and $\sin 2 \beta$ enables us to
evaluate the product of the neutralino masses through the determinant
equation \(neutdet). Then we can further bracket the neutralino
masses by invoking the near degeneracies with the chargino masses.

When $\mu$ is comparable to or greater than $\mglu$, then we see from
\(pertchar) that one of the charginos is lighter than the gluino,
and one is heavier. When $\mu$ is smaller than the gluino mass, then
applying the present bound on the gluino mass ($100$ GeV)
to eqs.~\(charsqsum) and  \(charprod), we see that one of the
charginos is still lighter than the gluino. Thus in all cases, at least one
chargino is lighter than the gluino.

By using a panoply of sum rules, some of which are new, we have been able to
analyze the qualitative features of the spectrum
of squarks, sleptons, charginos, and neutralinos.

\noindent{\bf Mass Orderings}

We repeat the main features of the spectrum:

\noindent $\bullet$ The squark and slepton spectrum is determined by
$m_0$, which sets the
overall scale, and $\mglu$ which sets their splitting into bands.

\noindent $\bullet$ The chargino and neutralino masses
are determined by $\mu$ and $\mglu$.

Thus it is fortunate that,
because of its strong interactions, it is quite likely that the gluino
will be the first sparticle to be found. Below we assume knowledge of $\mglu$
and proceed to discuss several possibilities.

As we have seen, there is at least one chargino which is lighter than the
gluino. However, the lightest chargino may not be the lightest charged
sparticle. There is a competition between the lightest chargino and
the right-handed sleptons for the honor of being the lightest
charged supersymmetric (odd $R$-parity) particle. When $m_0$ is large,
the chargino certainly wins, but in the ``no-scale"-type models,
the answer is less clear and depends most crucially on the value
of the parameter $\mu$.

On the other hand, the relative value of the squark and gluon masses is not
determined, since it depends directly on the input parameter $m_0$;
if $m_0$ is greater than $\mglu$, the squarks are heavier, and if $m_0$
is less than roughly $.5\> \mglu$ (see eq.~\(cthreeestlib)),
the squarks are lighter than the gluinos.
In the strict ``no-scale" limit $m_0 = 0$, we find from \(cthreeest) that the
squark band is centered at a mass between $.8 \> \mglu$ and $.9 \> \mglu$.
We see from eq.~\(sqsl) that this is the lightest the squark band
can be relative to the gluino.
As we have discussed
earlier, a large $m_0$ implies more clumping between sleptons and squarks.
The parameter $m_0$ is determined independently if
the right-handed selectron is found at a relatively low mass.

We can summarize the relative positions of the lightest
chargino, the right-handed selectron, the main squark band, and the gluino.
For small $m_0$,
$$
\mmer , M_{\tilde C_1} < M_{\tilde q} < \mglu \>.
\qquad\qquad (m_0 \approx 0)
$$
For intermediate values of $m_0$,  the situations
$$
M_{\tilde C_1} < \mmer < M_{\tilde q} < \mglu
\qquad\qquad (m_0 < .5 \mglu)
$$
$$
M_{\tilde C_1} < \mmer < \mglu < M_{\tilde q}
\qquad\qquad (.5 \mglu < m_0 < \mglu)
$$
can occur. However, for large enough $m_0$, one chargino and the gluino
are lightest:
$$
M_{\tilde C_1} < \mglu < \mmer < M_{\tilde q} \>.
\qquad\qquad ( m_0 > \mglu)
$$

The lightest chargino and the right-handed selectron are both
fine candidates to be pair-produced and studied at an $e^+e^-$
collider like the NLC or LEPII if they are light enough.
The chargino mass spectrum  depends on the parameters $\mu$ and
$\sin (2 \beta)$, as well as on the gaugino mass parameter $M_2$.
However, in our scenario for which the gluino is discovered
and well studied at a hadron collider, the value of $M_2$ follows
from knowledge of the gluino mass and $\alpha_3$ at that scale.
Then, knowledge of the chargino masses allows us to determine
$\mu$ and $\sin (2\beta)$. From these two parameters one can
in principle derive the whole neutralino spectrum as well,
since $M_1$ is also known once we measure the gluino mass.
In the end, the consistency of this picture becomes a numerical
question of putting constraints on the input parameters of
the theory through equations \(charsqsum) and \(charprod)
for the charginos and \(neutdet) and \(neutsqsum) for the neutralinos.

We thank W.~Bardeen and the SSC theory group
and also the Physics Department of
Southern Methodist University for their warm hospitality during the completion
of this work. We also wish to thank R.~Arnowitt, D.~Casta\~no,
G.~Kane, and S.~Meshkov
for useful discussions.

\references

\refis{reviews}
For reviews, see H.~P.~Nilles,  \prpts 110, 1, 1984 and
H.~E.~Haber and G.~L.~Kane, \prpts 117, 75, 1985.

\refis{supergrav}
A.~Chamseddine, R.~Arnowitt and P.~Nath,
\journal Phys.~Rev.~Lett., 49, 970, 1982;
H.~P.~Nilles,
\journal Phys.~Lett., 115B, 193, 1982;
L.~E.~Ib\'a\~nez,
\journal Phys.~Lett., 118B, 73, 1982;
R.~Barbieri, S.~Ferrara, and C.~Savoy,
\journal Phys.~Lett., 119B, 343, 1982;
L. Hall, J. Lykken and S. Weinberg,
\journal Phys. Rev., D27, 2359, 1983;
P. Nath, R. Arnowitt and A. H. Chamseddine,
\journal Nuc. Phys., B227, 121, 1983.

\refis{JLC}
See for instance, KEK Report 92-16 (JLC-I), and references
contained therein.

\refis{numer}
G.~G.~Ross and R.~G.~Roberts,
\journal Nuc.~Phys., B377, 571, 1991.

\refis{AN}
R. Arnowitt and P. Nath,
SSCL-Preprint-229;
\prl 69, 725, 1992;
\pr D46, 3981, 1992.

\refis{AN2}
R. Arnowitt and P. Nath,
\pl B289, 368, 1992.

\refis{num}
L.~E.~Ib\'a\~nez and  G.~G.~Ross,
``Electroweak Breaking in Supersymmetric Models",
CERN-TH.6412/92, in Perspectives in Higgs Physics,
G.~Kane, editor.

\refis{nume}
P.~Ramond, ``Renormalization Group Study of the Minimal Supersymmetric
Standard Model: No Scale Models", invited talk at workshop ``Recent
Advances in the Superworld", Houston TX, April 1993.

\refis{sla}
D.~J.~Casta\~no, E.~J.~Piard, and P.~Ramond, to appear.

\refis{trigger}
L.~E.~Ib\'a\~nez and  G.~G.~Ross,
\journal Phys.~Lett., 110B, 215, 1982;
K.~Inoue, A.~Kakuto, H.~Komatsu, and S.~Takeshita,
\journal Prog. Theor. Phys., 68, 927, 1982;
L.~Alvarez-Gaum\'e, M.~Claudson, and M.~Wise,
\np B207, 16, 1982;
J.~Ellis, J.~S.~Hagelin, D.~V.~Nanopoulos, and
K.~Tamvakis,
\journal Phys.~Lett., 125B, 275, 1983.

\refis{spartnersa}
L.~Alvarez-Gaum\'e, J.~Polchinski, and M.~Wise,
\np B221, 495, 1983;
L.~E.~Ib\'a\~nez, C.~L\'opez, and C.~Mu\~noz,
\np B256, 218, 1985;

\refis{spartnersb}
J.~S.~Hagelin and S.~Kelley,
\np B342, 95, 1990; A.~E.~Faraggi, J.~S.~Hagelin, S.~Kelley, and
D.~V.~Nanopoulos,
\pr D45, 3272, 1992.

\refis{sscfgrp} S.~P.~Martin \prd 46, 2769, 1992.

\refis{rparity}  A.~Font, L.~E.~Ib\'a\~nez and F.~Quevedo,
\pl B228, 79, 1989;
L.~E.~Ib\'a\~nez and  G.~G.~Ross, \pl B260, 291, 1991, \np B368, 3, 1992.

\refis{unification}
U.~Amaldi, W.~de Boer, and H.~Furstenau,
\pl B260, 447, 1991;
J.~Ellis, S.~Kelley and D.~Nanopoulos,
\pl 260B, 131, 1991;
P.~Langacker and M.~Luo,
\pr D44, 817, 1991.

\refis{GUT}
J.~C.~Pati and A.~Salam,
\pr D10, 275, 1974;
H.~Georgi and S.~Glashow,
\prl 32, 438, 1974;
H.~Georgi, in {\it Particles and Fields-1974}, edited by C.E.Carlson,
AIP Conference Proceedings No.~23 (American Institute of Physics,
New York, 1975) p.575;
H.~Fritzsch and P.~Minkowski,
\journal Ann.~Phys.~NY, 93, 193, 1975;
F.~Gursey, P.~Ramond, and P.~Sikivie,
\pl 60B, 177, 1975.

\refis{char} P.~Nath, A. Chamseddine, and R.~Arnowitt,
``Applied $N=1$ Supergravity", ICTP
Series-Vol. 1 (World Scientific 1983).

\refis{btau} H.~Arason, D.~J.~Casta\~no, B.~Keszthelyi, S.~Mikaelian,
E.~J.~Piard, P.~Ramond, and B.~D.~Wright,
\prl 67, 2933, 1991;
A.~Giveon, L.~J.~Hall, and U.~Sarid,
\pl 271B, 138, 1991.

\refis{lighthiggs} See for instance G.~L.~Kane, C.~Kolda, and J.~D.~Wells,
``Calculable Upper Limit on the Mass of the Lightest Higgs Boson in Any
Perturbatively Valid Supersymmetric Theory", Michigan preprint UM-TH-93-24,
to appear in Phys.~Rev.~Lett.,
and references therein.

\endreferences\end